\documentstyle[prl,aps,epsf,multicol]{revtex}

\begin{document}
\draft

\title{Test of molecular mode coupling theory: A first resume}
\author{C. Theis$^{1}$, F. Sciortino$^{2}$, A. Latz$^{3}$, R. Schilling$^{3}$ and P. Tartaglia$^{2}$}
\address{$^{1}$ Fakult\"at f\"ur Physik, Albert--Ludwigs--Universit\"at,
Hermann--Herder--Stra{\ss}e 3, D--79104 Freiburg, Germany \\
$^{2}$ Dipartimento di Fisica and Istituto Nazionale per la Fisica
della Materia, Universit\'{a} di Roma "La Sapienza", Piazzale Aldo
Moro 2, I--00185 Roma, Italy \\ $^{3}$ Institut f\"ur Physik,
Johannes Gutenberg Universit\"at, Staudinger Weg 7, D--55099
Mainz, Germany}

\date{\today}
\maketitle

\begin{abstract}
We report recent progress on the test of mode coupling theory for
molecular liquids (MMCT) for molecules of arbitrary shape. The
MMCT equations in the long time limit are solved for supercooled water 
including all
molecular degrees of freedom. In contrast to our earlier treatment of
water as a linear molecule, we find that the glass transition
temperature $T_c$ is overestimated by the theory as was found in
the case of simple liquids. The nonergodicity parameters
are calculated from the "full" set of MMCT-equations truncated at $l_{co}=2$.
These results are compared $(i)$ with the nonergodicity parameters from MMCT
with $l_{co}=2$ in the "dipole" approximation $n=n'=0$ and the diagonalization
approximation $n=n'=0, \, l=l'$ and $(ii)$ with the corresponding results from
a MD-simulation. This work supports the possibility that a reduction to the 
most prominent correlators may constitute a valid approximation for solving
the MMCT equations for rigid molecules.
\end{abstract}


\begin{multicols}{2} 
\section{Introduction}
\label{sec:intro}
The mode coupling theory (MCT) for supercooled simple
liquids proposed by Bengtzelius, G\"otze and Sj\"olander
\cite{bengt84} 
interprets the glass transition
as a dynamical transition. 
This picture has been supported by many 
experiments on several glass formers  (see e.g.
\cite{gotze99} and Refs. therein) and more recently by detailed analysis of
computer simulations for
Lennard-Jones systems (see e.g. \cite{nauro97} and references therein). 
The signature of the
dynamical transition, i.e. the asymptotical power laws, have been
discovered also via neutron scattering, light scattering, dielectric
relaxation and NMR in {\em molecular} glassformers (see
\cite{gotze99} and references therein), stimulating the
extension of MCT to molecular liquids.

Two approaches for such an extension have been proposed recently for rigid
molecules.
Chong and Hirata \cite{hirata} introduced a theory based
on a site-site description of the molecules. Their approach offers
the advantage to be closely related to neutron scattering
experiments. The structural information --- which is a necessary
input of the theory --- can be readily obtained from  theories of 
molecular liquids able to predict partial structure factors, 
like the RISM \cite{hanse86} approximation. 
The second approach 
is based on the expansion of the orientational density
into a complete set of functions, in analogy to
the Fourier expansion of the density related to the translational
degrees of freedom.  
To distinguish the second approach from the site-site theory,  it is called
molecular mode coupling theory (MMCT).
MMCT was derived for a single linear
molecule in a simple liquid \cite{frano97}, for liquids of linear
molecules \cite{schil97}(for some application see \cite{letz}) and for
molecules of  arbitrary
shape \cite{kawas97,fabbi99}. MMCT allows to calculate the
glassy dynamics for all
orientational degrees of freedom, and  it is closely connected to dielectric
and NMR experiments. As reorientational motion is
also very important in light scattering experiments \cite{cummi,murry99}
it may also be helpful in their interpretation. 
The connection of MMCT 
to neutron scattering experiments has been discussed recently
\cite{theis99}. 

The fundamental MMCT quantities are the time-dependent correlation
functions 
\begin{equation}
S_{ln,l'n'}(q,m,t)=\langle \rho_{ln}^\ast(q,m,t) \rho_{l'n'}(q,m)
\rangle \label{eq:sqt}
\end{equation}
of the tensorial density modes
\begin{equation}
\rho_{ln}(q,m,t)=i^l(2l+1)^\frac{1}{2}\sum_{j=1}^N e^{i \vec{q}
\vec{x}_j(t)} {\cal D}_{mn}^{l \ast}(\Omega_j(t)).
\label{eq:modes}
\end{equation}
Here it is ${\bf q}=(0,0,q)$ and $l$ runs over all positive integers 
including zero, $m$ and
$n$ take integer values between $-l$ and $l$, ${\cal D}$ denotes
the Wigner functions \cite{gray84}. The reader should note that the
correlators (\ref{eq:sqt}) are diagonal in $m$ for ${\bf q}=(0,0,q)$.
The MMCT equations of motion for the
Laplace transform ${\bf S}(q,m,z)=i \int_0^\infty {\bf S}(q,m,t)
e^{izt}$, $(\mbox{Im } z > 0)$ have been presented in 
a preceding paper.  Here we focus on the {\it unnormalized}
molecular nonergodicity parameters
\begin{equation}
{\bf F}(q,m)=\lim_{t \to\infty} {\bf S}(q,m,t)=-\lim_{z \to 0}
z {\bf S}(q,m,z) \label{eq:nonerg}
\end{equation}
which we calculated using the following set of equations
\cite{fabbi99}.
\begin{eqnarray}
& &{\bf F}(q,m) = \\ 
& & \quad \left[ {\bf S}^{-1}(q,m) + {\bf S}^{-1}(q,m)
{\bf K}(q,m) {\bf S}^{-1}(q,m) \right]^{-1} \label{eq:nonerg1} \nonumber 
\end{eqnarray}
\end{multicols}
\twocolumn
\begin{eqnarray}
& & K_{ln, l'n'}(q,m) = \\
& & \quad \sum_{\alpha \alpha'} \sum_{\mu \mu'}
q^{\alpha \mu }_{ln}(q) q^{\alpha' \mu' \ast}_{l'n'}(q) \left(
{\bf \underline{m}}^{-1}(q) \right)^{\alpha \mu, \alpha' \mu'}_{lmn,
l'mn'} \label{eq:nonerg2} \\
& & m^{\alpha \mu, \alpha'
\mu'}_{lmn, l'm'n'}(q) = \frac{\rho_0}{(8 \pi^2)^3} \int\limits_0^\infty
dq_1 \int\limits_{|q-q_1|}^{q+q_1} dq_2 \nonumber \\
& & \quad \sum_{m_1 m_2} \sum_{{l_1 l'_1 \atop l_2
l'_2}} \sum_{{n_1 n'_1 \atop n_2 n'_2}} 
\, v^{\alpha \mu}_{ln, l_1 n_1, l_2 n_2}(q q_1
q_2; m m_1 m_2) \nonumber \\
& & \quad v^{\alpha' \mu' \ast}_{l'n', l'_1 n'_1, l'_2
n'_2}(q q_1 q_2; m' m_1 m_2) \; F_{l_1 n_1, l'_1
n'_1}(q_1,m_1) \, \nonumber \\ & & \quad F_{l_2 n_2, l'_2 n'_2}(q_2,m_2)
\label{eq:nonerg3}
\end{eqnarray}

The memory function matrix $\underline{{\bf m}}$ in
Eq.(\ref{eq:nonerg3}) represent the mode coupling approximation
for the correlation function of fluctuating forces. 
The
index $\alpha$ labels the 
translational ($\alpha = T$) and rotational ($\alpha = R$) currents,
each of them consisting of three (spherical) vector components $\mu \in
\{-1,0,1\}$. The vertex functions $v$ are determined only by the
matrix of the static molecular structure factors ${\bf S}(q,m)$ and the number
density $\rho_0$. Their explicit form has been given in Ref.
\cite{fabbi99}. 
The coefficients $q_{ln}^{\alpha \mu}(q)$
appearing in Eq.(\ref{eq:nonerg2}) are 
\begin{equation}
q_{ln}^{\alpha \mu}(q)= \left\{
\begin{array}{c@{\quad}c}
0 & \alpha = T, \mu = \pm 1 \\ q & \alpha = T, \mu = 0 \\
\frac{1}{\sqrt{2}} \sqrt{l(l+1)-n(n+\mu)} & \alpha = R, \mu \pm 1
\\ n & \alpha = R, \mu = 0 \\
\end{array}
\right. .
\label{eq:qcoeffs}
\end{equation}
Since $q_{ln}^{T \pm 1}=0$ (due to the choice of ${\bf q}=(0,0,q)$)
the transversal translational
components ($\alpha=T \, , \, \mu= \pm 1$) of the memory functions
enter only indirectly (via the inversion of $\underline{{\bf
m}}(q)$) and thus shall be neglected in the following.

The given set of equations (\ref{eq:nonerg1})-(\ref{eq:nonerg3})
includes all interactions between translational and
rotational degrees of freedom in molecular liquids and thus
this set is rather involved.
Obviously its numerical solution poses a formidable task.

As model system for our analysis we have chosen SPC/E water
\cite{beren87}. The water molecule possesses a twofold rotational
symmetry ($C_{2v}$) around the axis given by its dipole moment
which has been chosen as the z-axis of the body fixed frame. 

\begin{table}[b]
\begin{tabular}{|c|c|c|c|c|}
\hline  & MD & MMCT diag & MMCT dipole & MMCT "full" \\ \hline $T_c$ & 200
K & 206 K & 208 K & 279 K \\ \hline
\end{tabular}
\caption{Comparison of the critical temperatures $T_c$ for the
MD-simulation and the theoretical calculations in diagonal-dipole
approximation, dipole-approximation and for the "full" theory.}
\label{tab:Tc}
\end{table}

The
z-axis and the x-axis define the plane which is spanned by the
molecule. As discussed in detail in Ref. \cite{fabbi99} the
$C_{2v}$-symmetry can be used to simplify the equations of motion
by the restriction that $n$ and $n'$ are even. 
Preceeding
publications on long time MD--simulations \cite{sciortino} for
this strong glass former showed that the centre of mass dynamics
is in good agreement with the predictions of the asymptotic laws
of MCT. Recent work \cite{fabbi98a} demonstrated that the
signature of the dynamic transition can also be observed for the
orientational degrees of freedom of the molecule.

A first approach to solve Eqs.
(\ref{eq:nonerg1})-(\ref{eq:nonerg3}) has been given in Ref. \cite{fabbi99}. 
Apart from the necessary
truncation of the range of $l$ by a cutoff $l_{co}$ for which we
have chosen $l_{co}=2$ we introduced in \cite{fabbi99} two more 
approximations. In the {\it dipole approximation}, water was
treated as a linear molecule oriented in the direction of its
dipole moment. This approximation corresponds to set 
in Eqs. (\ref{eq:nonerg1})-(\ref{eq:nonerg3}) 
the angular indices $n$, $n'$ and the corresponding summation indices
$n_i$,$n_i'$ to zero. 
In Ref. \cite{fabbi99} we have also studied the stronger 
{\em diagonal-dipole approximation}, which
makes the further assumption that structure factors, nonergodicity
parameters and memory functions are diagonal with respect to the
angular indices $l$ and $l'$. It is the main purpose of the present 
communication to
present results obtained by solving the {\em "full"} set of
MMCT-equations up to $l_{co}=2$ without any further approximation.
The main difference to the approach in ref. \cite{fabbi99} is that we take the 
nonlinear character of the water molecule serious. In particular, this means 
that the rotational motion of both protons around the dipolar axis is taken 
into account. Therefore, we can study the influence of this degree of freedom
on the ideal glass transition temperature $T_c$ and on the nonergodicity
parameters $F_{l 0,l' 0}(q,m)$ obtained in ref. \cite{fabbi99}. In addition we
also obtain the new parameters $F_{ln,l'n'}(q,m)$ with $n$ and/or $n'$
different from zero.

\section{Results}
\label{sec:results}
We have solved  Eqs.
(\ref{eq:nonerg1})-(\ref{eq:nonerg3}) iteratively on a grid of
100 equispaced wave-vectors, up to 110.7 $nm^{-1}$.
One complete iteration requires 4 days of cpu time on a
533 MHz  alpha workstation. The entire calculation to locate the
critical temperature and the corresponding nonergodicity
parameter, with a tolerance of $2.5 \times 10^{-4}$ per point, 
requested more than
250 iterations. On a dedicated 4-nodes parallel machine it required about 250 days.  
The number of iteration at the critical temperature was $54$.
We found that at $T=282 K$ the MMCT equations predict a liquid phase,
while at $T=272 K$ unambiguously a glassy
state is predicted. Within the chosen tolerance, we locate $T_c^{MCT}$ 
at $T_c^{MCT}=279 K$.

Table \ref{tab:Tc} summarizes the critical temperatures at which a
transition from ergodic to nonergodic behavior is found. 
While
$T_c$ is almost equal for both approximations used in ref. \cite{fabbi99}
and quite close to
the result of the MD-simulation we find that the critical
temperature is overestimated by almost 50\% using the "full" theory.
Thus we confirm our supposition in \cite{fabbi99} that the agreement for $T_c$ 
between simulation and both 
approximation schemes was fortuitous. Such finding could also be
expected from the static structure factors used as input.
Besides the dominating diagonal structure factors
$S_{00,00}(q,m=0)$ and $S_{10,10}(q,m)$ the most prominent peaks
are displayed by $S_{10,2 \pm 2}(q,m)$ and $S_{2 \pm 2,2 \pm
2}(q,m)$ (See Figures in Ref. \cite{fabbi98b}), which were neglected
in \cite{fabbi99}.
The overestimation of the critical temperature is common
to the MCT for simple liquids \cite{nauro97}
and seems to be a general deficiency of the mode coupling
approximation. Thus we see that --- although the overestimation of the
critical temperature is not desired --- it is necessary to include all
static structure factors with large amplitude to get a concise description
in the MMCT framework.

\begin{figure}[htb]
\centerline{\epsfxsize=8cm \epsfysize=7cm
\epsffile{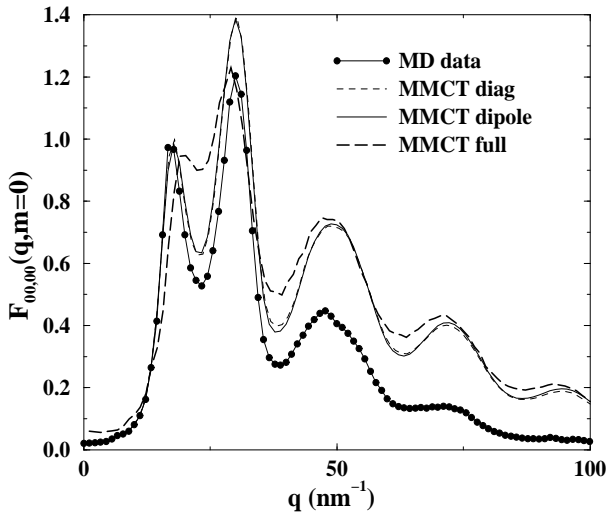}} \caption{Critical nonergodicity
parameters $F_{00,00}(q,m=0)$ for the centre of mass.
The results from the MD simulation (symbols) are compared with the theoretical 
predictions of MMCT in diagonal-dipole approximation (short dashed line), 
in dipole approximation (solid line) and the results obtained using the "full"
set of MMCT-equations (long dashed line).} \label{fig:COM}
\end{figure}

Figure \ref{fig:COM} shows the comparison of the {\em unnormalized} critical
nonergodicity parameters $F_{00,00}(q,m=0)$ for the centre of mass
correlations. The oscillations of the MD-result are captured well by
all of our 
MMCT-results. As for the critical temperature we can also observe for the
critical nonergodicity parameters that the dipole approximation and the
diagonal-dipole approximation lead to nearly the same result.
In the vicinity of the maximum of the structure
factor the agreement between theory and simulation is improved by
removing the additional approximations but for the prepeak as well
as for large $q$ both approximation schemes perform
better than the "full" theory. In the region of the prepeak and especially for
the minimum between prepeak and main peak we observe that the oscillations are
less pronounced and the peak positions are slightly shifted. Exactly the same
behaviour can be found by a comparison of the static structure factors at the
different critical temperatures. Thus the worse performance of the "full" theory
in comparison with the approximation schemes can be attributed at least partly
to the overestimation of the critical temperature because of which the static
input of the calculations does not reflect accurately the static structure at
the "true" $T_c$ of the simulation. Apart from that, it also has to be taken
into account that in spite of the computational effort we have made, the fixed
point of Eq. (\ref{eq:nonerg1})-(\ref{eq:nonerg3}) can not be determined with
very high precision. Therefore, the result for the critical nonergodicity
parameters after 54 iterations may overestimate the exact one by a few
percent. The latter reason may also be responsible for the worse performance
of the "full" theory at the minima of $F_{00,00}(q,m=0)$ as Nauroth
\cite{naurothdipl} found that the convergence of the iteration is extremely
slow in these parts. Finally we want to mention that the disagreement at large
$q$ between simulation and {\em all} theoretical calculations was also found
for simple liquids \cite{nauro97} and was considered as a shortcoming of the
mode coupling approximation. 

\begin{figure}[htb]
\centerline{\epsfxsize=8cm \epsfysize=8cm
\epsffile{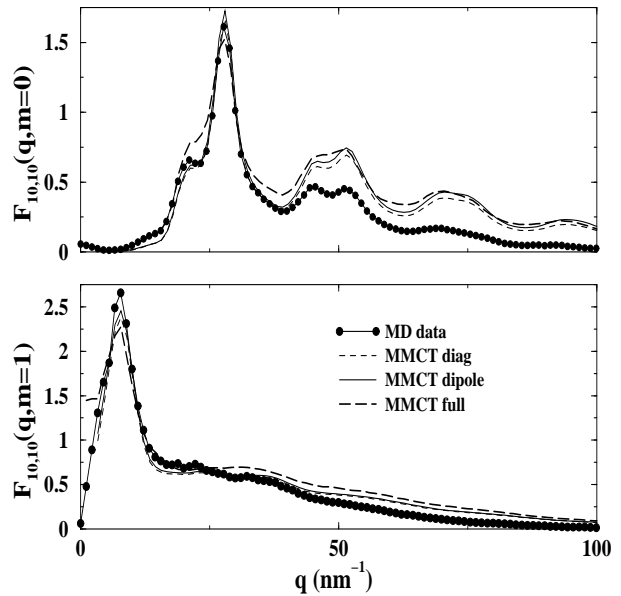}} \caption{Diagonal critical nonergodicity
parameters $F_{10,10}(q,m)$ as  calculated from the MD simulation
(symbols) compared with the theoretical predictions of MMCT in
diagonal-dipole approximation (short dashed line), in dipole
approximation (solid line) and the results obtained using the "full"
set of MMCT-equations (long dashed line).} \label{fig:Fmx10xp10}
\end{figure}

As in the case of the centre of mass correlators the
$q$-dependence of the nonergodicity parameters $F_{10,10}(q,m)$
(see fig. \ref{fig:Fmx10xp10}) is reproduced well by the theory.
Here even in the vicinity of the maximum the approximations
perform better than the "full" theory. 

\begin{figure}
\centerline{\epsfxsize=8cm \epsfysize=9cm
\epsffile{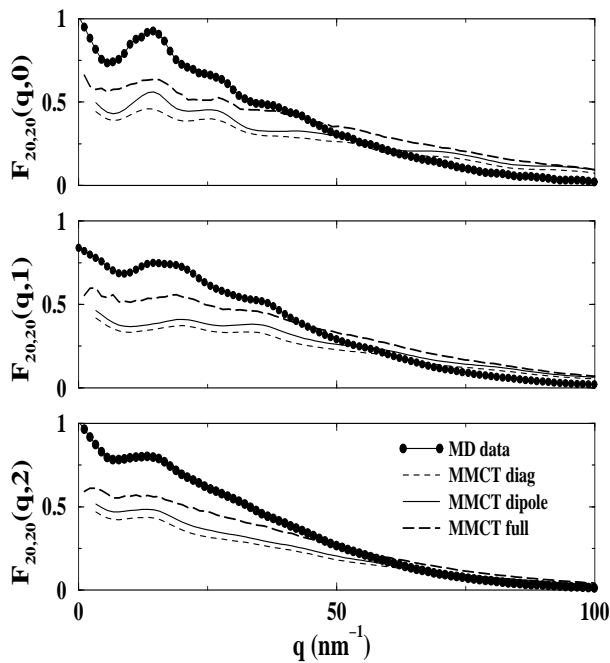}} \caption{Diagonal critical nonergodicity
parameters $F_{20,20}(q,m)$ as calculated from the MD simulation
(symbols) compared with the theoretical predictions of MMCT in
diagonal-dipole approximation (short dashed line), in dipole
approximation (solid line) and the results obtained using the "full"
set of MMCT-equations (long dashed line).} \label{fig:Fmx20xp20}
\end{figure}

Figure \ref{fig:Fmx20xp20} shows the comparison for the critical
nonergodicity parameters $F_{20,20}(q,m)$. Apart from the region
of large wave vectors the "full" theory shows better agreement with
the simulation than the two approximations. Thus one observes that
the approximations, which neglect terms with $n\not=0$ and
consequently $l \ge 2$, have stronger effect on the
$l=2$-correlators than on those for $l=1$ or $l=0$. 
Further one observes that the agreement between theory and simulation is less
good for the $l=2$--correlators than for those with $l=0$ or $l=1$. The reader
should note that the good agreement at large $q$ for the $l=2$--correlators
must be considered as fortuitous because the overestimation of the
nonergodicity parameters (seen for $l=0,1$) is compensated by a generally too
small amplitude of the $l=2$--correlators. 

\noindent The reason for the worse agreement
of the $l=2$--correlators is their higher sensibility to the cutoff at
$l_{co}=2$ which can
be understood on a mathematical level by a closer examination of the vertices
which have been given in ref. \cite{fabbi99}. 
Let us pick out one special
example to illustrate this point. The vertex factor 
$v^{\alpha \mu}_{00,20,20}(q,q_1,q_2;0,0,0)$ is responsible for the coupling
of two correlators involving $l=0$ and  $l=2$, respectively. It is of the form
\begin{eqnarray} \label{vert02}
& &v^{\alpha \mu}_{00,20,20}(q,q_1,q_2;0,0,0)= \\
& & \quad \sum_{l,n} u^{\alpha
\mu}_{00,ln,20}(q,q_1,q_2;0,0,0) c_{ln,20}(q_1,0) + (1 \leftrightarrow 2), 
\nonumber
\end{eqnarray}
where $u^{\alpha \mu}_{00,ln,20}(q,q_1,q_2;0,0,0)$ contains 
Clebsch-Gordan coefficients of the form ${\cal C}(l,2,0;m',-m',0)$
which 
enforces $l=2$. 
Therefore the contribution to the memory function matrix
caused by the vertex factor in Eq. (\ref{vert02}) is
exact even for $l_{co}=2$. 
The vertex factor $v^{\alpha \mu}_{20,20,20}(q,q_1,q_2;0,0,0)$ instead, 
which describes the 
coupling of two correlators involving $l=2$ and  $l=2$,
respectively, has the form
\begin{eqnarray}
& &v^{\alpha \mu}_{20,20,20}(q,q_1,q_2;0,0,0)= \\
& & \quad \sum_{l,n} u^{\alpha
\mu}_{20,ln,20}(q,q_1,q_2;0,0,0) c_{ln,20}(q_1,0) + (1 \leftrightarrow 2). 
\nonumber
\end{eqnarray}
The corresponding Clebsch-Gordan coefficient \linebreak ${\cal C}(l,2,2;m',-m',0)$ allows
$l\in\{0,1,2,3,4\}$. Thus for $l_{co}=2$ this vertex is not entirely 
taken into acount, 
although it is responsible for the coupling of $l=2$-correlations. 
From this example we
can see that the $l=2$-quantities are more sensitive to the cutoff than those
for $l=0$. 

\begin{figure}[htb]
\centerline{\epsfxsize=8cm \epsfysize=8cm \epsffile{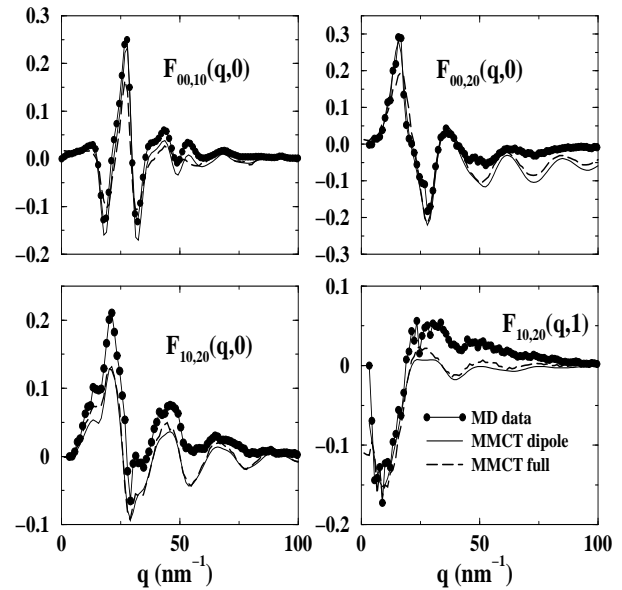}}
\caption{Off diagonal critical nonergodicity parameters
$F_{l0,l'0}(q,m)$ with $l\not=l'$ as calculated from the MD
simulation (symbols) compared with the theoretical predictions of
MMCT in dipole approximation (solid line) and the results obtained
using the "full" set of MMCT-equations (long dashed line).}
\label{fig:Fdipol}
\end{figure}

The {\em off diagonal} critical nonergodicity parameters with $n=n'=0$ shown in
figure \ref{fig:Fdipol} are reproduced well by the theory. The
difference between dipole-approximation and "full" theory are
relatively small.

\begin{figure}[htb]
\centerline{\epsfxsize=8cm \epsfysize=9cm \epsffile{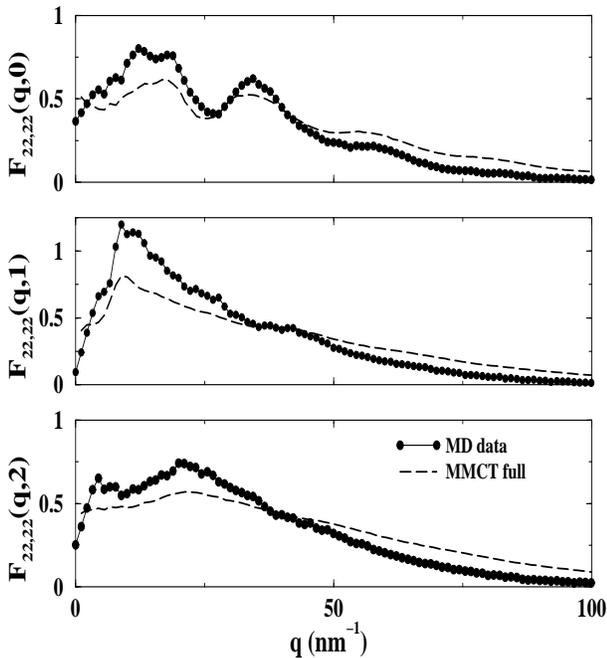}}
\caption{Diagonal critical nonergodicity parameters
$F_{22,22}(q,m)$ with $n\not=0$ as calculated from the MD
simulation (symbols) compared with the results obtained using the
"full" set of MMCT-equations (long dashed line).}
\label{fig:Fn2diag}
\end{figure}

\begin{figure}[htb]
\centerline{\epsfxsize=8cm \epsfysize=9cm
\epsffile{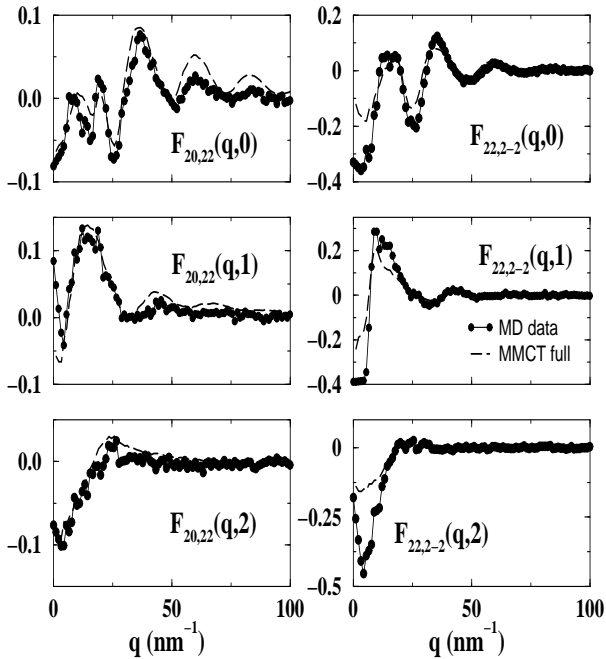}} \caption{Diagonal critical
nonergodicity parameters $F_{2n,2n'}(q,m)$ with $n\not=0$ and/or
$n'\not=0$ as calculated from the MD simulation (symbols) compared
with the results obtained using the "full" set of MMCT-equations
(long dashed line).} \label{fig:Fn2nondiag}
\end{figure}

Figures \ref{fig:Fn2diag} and \ref{fig:Fn2nondiag} show the
comparison between the new critical nonergodicity parameters with $n\not=0$
and/or $n'\not=0$ and the simulation results. Since those terms
\noindent are neglected by the approximations they can only be calculated
using the "full" set of MMCT-equations. The agreement between theory
and simulation is satisfactory for almost all correlators. 

In the discussion of the {\it unnormalized} critical nonergodicity
parameters we have seen that 
the worse performance of the "full" theory is partly due to the fact that the
static input has to be taken at the "wrong" temperature. This influences the
nonergodicity parameters in two ways: $i)$ ${\bf S}(q,m)$ is the initial value
of the correlation function ${\bf S}(q,m,t)$ whose limit for $t \to\infty$ are
the {\it unnormalized} nonergodicity parameters $F_{ln,l'n'}(q,m)$. 
$ii)$ In the mode coupling approximation the
static structure factors determine the vertices which describe the
coupling between 
the tensorial density modes in the system. This influence can partly
be eliminated by 
calculating the {\em normalized} critical nonergodicity parameters 
\begin{equation}
f_{ln,l'n'}(q,m)=\frac{F_{ln,l'n'}(q,m)}{\sqrt{S_{ln,ln}(q,m)
S_{l'n',l'n'}(q,m)}}. \label{eq:normed}
\end{equation}

\begin{figure}[htb]
\centerline{\epsfxsize=8cm \epsfysize=7cm \epsffile{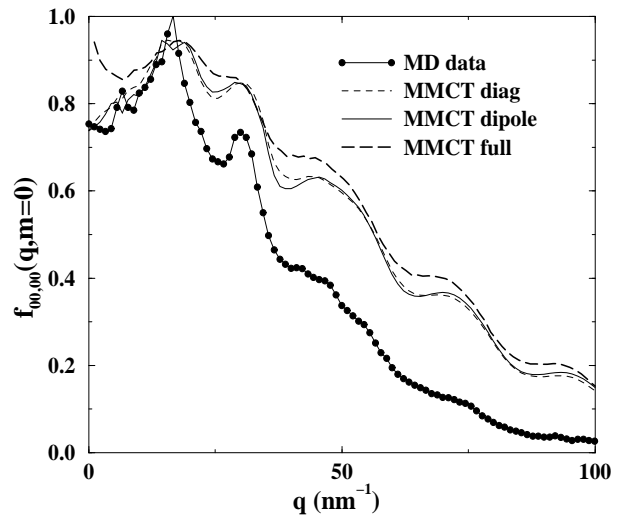}}
\caption{Normalized critical nonergodicity parameters
$f_{00,00}(q,0)$ for the centre of mass.
The results from the MD simulation (symbols) are compared with the theoretical 
predictions of MMCT in diagonal-dipole approximation (short dashed line), 
in dipole approximation (solid line) and the results obtained using the "full"
set of MMCT-equations (long dashed line).} \label{fig:fm0x00xp00}
\end{figure}

\begin{figure}[htb]
\centerline{\epsfxsize=8cm \epsfysize=8cm \epsffile{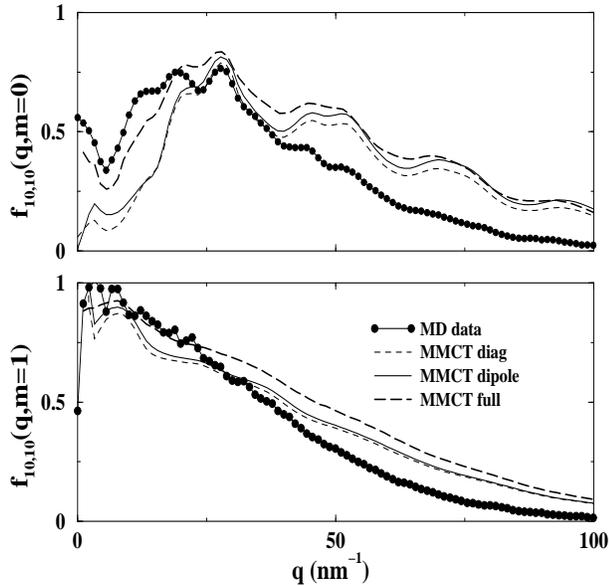}}
\caption{Diagonal normalized critical nonergodicity parameters
$f_{00,00}(q,m)$ as calculated from the 
MD simulation (symbols) compared with the theoretical 
predictions of MMCT in diagonal-dipole approximation (short dashed line), 
in dipole approximation (solid line) and the results obtained using the "full"
set of MMCT-equations (long dashed line).} \label{fig:fmx10xp10}
\end{figure}

\begin{figure}[htb]
\centerline{\epsfxsize=8cm \epsfysize=9cm \epsffile{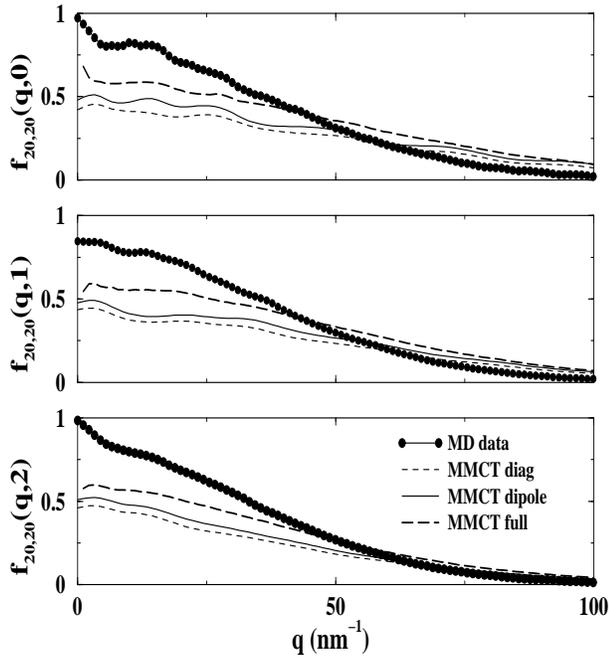}}
\caption{Diagonal normalized critical nonergodicity parameters
$f_{20,20}(q,m)$ as calculated
from the MD simulation (symbols) compared with the theoretical 
predictions of MMCT in diagonal-dipole approximation (short dashed line), 
in dipole approximation (solid line) and the results obtained using the "full"
set of MMCT-equations (long dashed line).} \label{fig:fmx20xp20}
\end{figure}

A selection of the normalized diagonal nonergodicity parameters is 
shown in Fig.(\ref{fig:fm0x00xp00})-(\ref{fig:fmx20xp20}). Due to the
normalization the variation with $q$ is less pronounced. We observe that the
difference in amplitude between the "full" theory and the approximations is
reduced as the initial value ${\bf S}(q,m,t=0)$ of the time-dependent
correlation functions is
set to 1 at all wavevectors and all temperatures. Nonetheless the dipole and
diagonal-dipole approximation still provide the better description of the
normalized critical nonergodicity parameter. 

\noindent In part this may be explained by the better 
convergence of the iteration for the approximation schemes, but the wrong
critical temperature also has its influence as can be seen from the fact that
the peak positions for the "full" MMCT results are still slightly shifted.

\section{Summary and Conclusions}
\label{sec:conclusion}

From our analysis we can state the following conclusions:
{\em i)} The mode coupling theory for molecular liquids
overestimates the critical temperature $T_c$ in the same fashion
as its counterpart for simple liquids. {\em ii)} The
$q$-dependence of the critical nonergodicity parameters is well
reproduced in the vicinity of the main peaks. Systematic
differences exist for large wave vectors. {\em iii)} Deviations
between theory and simulation are partly due to the overestimation
of the critical temperature. Therfore the structural input of
the theory does not reflect properly the structure of the liquid
at the true glass transition temperature. These deviations still exist 
for the normalized nonergodicity parameters. {\em iv)} Approximation
schemes taking into account only part of the correlators like the dipole and
the diagonal-dipole approximation can
already give a reasonable description. {\em v)} The essential correlation
functions can be selected on the basis of the static structure
factors. I.e. for supercooled water the most important static
correlator are those with $n\ne 0$ and/or $n'\ne 0$ and for $n=n'=0$
those, which are diagonal in $l$ and $l'$. 

We think that in combination with suitable approximations for the static
structure factors MMCT will be a useful tool in the understanding of the glass
transition in molecular liquids. The solution of the "full" set of MMCT
equations is still a formidable task with present day computers, but
approximation schemes can be constructed on the basis of the
importance of various static structure
factors. A reduction to the most prominent correlators and $q$-vectors offers
the possibility to construct {\em microscopically based} 
schematic models for the description of molecular liquids.

\section{acknowledgement} 
The numerical work would not have been possible without the
support of INFM-PAIS-C and MURST-PRIN98. 
F.S. and P.T. acknowledge support also from INFM-PRA. A.L. and R.S. are 
grateful for financial support from the SFB-262.
We also thank L. Fabbian for participating to the early stages of this
project and providing us with part of the input data.


\begin{references}

\bibitem{bengt84} U.Bengtzelius, W.G\"otze, and A.Sj\"olander,
J. Phys. C {\bf 17}, 5915 (1984).


\bibitem{gotze99} W.G\"otze, J. Phys.: Condens. Matter {\bf 11},A1 (1999).


\bibitem{nauro97} W.Kob and H.C.Andersen, Phys. Rev. Lett {\bf73},
1376; Phys. Rev. E {\bf 51}, 4626 (1995);
Phys. Rev. E {\bf 52}, 4134 (1995);
M.Nauroth and W.Kob, Phys. Rev. E {\bf 55}, 657
(1997).
\bibitem{hirata} S.-H.Chong and F.Hirata, Phys. Rev. E {\bf 57},
1691 (1998); Phys. Rev. E {\bf 58},
6188 (1998); Phys. Rev. E {\bf 58},
7296 (1998).

\bibitem{hanse86} J.P.Hansen and I.R.McDonald, {\em Theory of Simple Liquids},
(Accademic Press, London, 1986).

\bibitem{frano97} T.Franosch, M.Fuchs, W.G\"otze, M.R.Mayr, and
A.P.Singh, Phys. Rev. E {\bf 56}, 5695 (1997).

\bibitem{schil97} R.Schilling and T.Scheidsteger, Phys. Rev. E
{\bf 56}, 2932 (1997).

\bibitem{letz} M.Letz and A.Latz, Phys. Rev. E {\bf 60}, 5865
(1999); M.Letz, R.Schilling, and A.Latz, cond-mat/9906336
(1999); M.Letz and R.Schilling, cond-mat/9908295 (1999).

\bibitem{kawas97} K.Kawasaki, Physica {\bf A 243}, 25 (1997).

\bibitem{fabbi99} L.Fabbian, A.Latz, R.Schilling, F.Sciortino,
P.Tartaglia, and C.Theis, Phys. Rev. E {\bf 60}, 5768 (1999).

\bibitem{cummi} H.Z.Cummins, G.Li, W.Du, R.M.Pick, 
and C.Dreyfus, Phys. Rev. E {\bf 53}, 896 (1996). \\ H.Z.Cummins, G.Li, W.Du, 
R.M.Pick, and C.Dreyfus, Phys. Rev. E {\bf 55}, 1232 (1997).

\bibitem{murry99} R.L.Murry, J.T.Fourkas, W.-X.Li, and T.Keyes,
Phys. Rev. Lett. {\bf 83}, 3550 (1999). 

\bibitem{theis99} C.Theis and R.Schilling, Phys. Rev. E {\bf 60},
740 (1999).

\bibitem{gray84} C.G.Gray and K.E.Gubbins, {\em Theory of
Molecular Fluids}, (Clarendon Press, Oxford, 1984), Vol. 1.

\bibitem{beren87} H.J.C.Berendsen, J.R.Grigera, and T.P.Straatsma,
J.Phys.Chem. {\bf 91}, 6269 (1987).

\bibitem{sciortino} F.Sciortino, L.Fabbian, S.-H.Chen, and
P.Tartaglia, Phys. Rev. E {\bf 56}, 5397 (1997); F.Sciortino,
P.Gallo, R.Tartaglia, and S.-H.Chen, Phys. Rev. E {\bf 54}, 6331
(1996); P.Gallo, F.Sciortino, P.Tartaglia, and S.-H.Chen, Phys.
Rev. Lett. {\bf 76}, 2730 (1996); F.W. Starr, S. Harrington, 
F. Sciortino and H.E. Stanley  Phys. Rev. Lett. {\bf 82}, 3629
(1999).

\bibitem{fabbi98a}  L.Fabbian, F.Sciortino, and P.Tartaglia, J. Non-Cryst.
Solids {\bf 235-237}, 350 (1998).

\bibitem{fabbi98b} L.Fabbian, A.Latz, R.Schilling, F.Sciortino,
P.Tartaglia, and C.Theis, cond-mat/9812363 (1998).



\bibitem{naurothdipl} M.Nauroth, Diploma thesis, Johannes-Gutenberg
Universit\"at Mainz, (1995).

\end{references}
\end{document}